# Electric Field and Tip Geometry Effects on Dielectrophoretic Growth of Carbon Nanotube Nanofibrils on Scanning Probes

Haoyan Wei[1,3], Anna Craig[1], Bryan D. Huey[1], Fotios Papadimitrakopoulos[2] and Harris L. Marcus[1,4]

[1] Materials Science and Engineering Program, Department of Chemical, Materials and Biomolecular Engineering, Institute of Materials Science, University of Connecticut, Storrs, CT 06269, USA

[2] Nanomaterials Optoelectronics Laboratory, Polymer Program, Institute of Materials Science, Department of Chemistry, University of Connecticut, Storrs, CT 06269, USA

[3] Present address: Applied Sciences Laboratory, Institute for Shock Physics, Washington State University, Spokane, WA 99210, USA, email: hwei@wsu.edu

Email: Haoyan Wei, wtaurus@msn.com

**Abstract**

Single-wall carbon nanotube (SWNT) nanofibrils were assembled onto a variety of conductive scanning probes including atomic force microscope (AFM) tips and scanning tunnelling microscope (STM) needles using positive dielectrophoresis (DEP). The magnitude of the applied electric field was varied in the range of 1-20V to investigate its effect on the dimensions of the assembled SWNT nanofibrils. Both length and diameter grew asymptotically as voltage increased from 5-18V. Below 4V, stable attachment of SWNT nanofibrils could not be achieved due to the relatively weak DEP force *versus* Brownian motion. At voltages of 20V and higher, low quality nanofibrils resulted from incorporating large amounts of impurities. For intermediate voltages, optimal nanofibrils were achieved, though pivotal to this assembly is the wetting behaviour upon tip immersion in the SWNT suspension drop. This process was monitored *in situ* to correlate wetting angle and probe geometry (cone angles and tip height), revealing that probes with narrow cone angles and long shanks are optimal. It is proposed that this results from less wetting of the probe apex, and therefore reduced capillary forces and especially force transients during the nanofibril drawing process. Relatively rigid probes (force constant $\geq$ 2N/m) exhibited no perceivable cantilever bending upon wetting and dewetting, resulting in the most stable process control.

PACS codes: 61.46.Fg, 87.64.Dz, 87.50.ch

**Keywords**

Single-wall carbon nanotubes, positive dielectrophoresis, scanning probe microscope, nanofibrils

## 1. Introduction

Following the invention of scanning tunnelling microscopy (STM) in 1981, a variety of related analytical methods have been developed in response to the increasing demand in nanoscale and atomic characterization, resulting in the measurement category collectively known as scanning probe microscopy (SPM). Atomic force microscopy (AFM), due to its capability to simultaneously characterize various surface topological, chemical, mechanical and electrical properties [1-3], and its versatility to function in different environments (*i.e.* vacuum, ambient, fluid, etc.), has quickly become the predominant SPM-based technique. In the process of scanning, a sharp Si or $Si_3N_4$ tip with a pyramidal shape and a typical cone angle of 20-30 degrees [4] is raster-scanned across a sample surface and various force interactions between the tip and sample are recorded by the probe. Although providing excellent spatial resolution, it is nevertheless limited by the tip radius sharpness and the aspect-ratio [5, 6]. Despite numerous efforts to reduce tip radii and increase aspect-ratios, tip-to-tip variation [4] and tip wear [7] continue to be major barriers to further improving conventional cone-shaped semiconductor-based AFM probes. One-dimensional carbon nanotubes (CNTs), owing to their high-aspect-ratio, small diameter, high stiffness, and buckle flexing capabilities, have thus emerged as a promising candidate for developing next generation AFM probes [5, 7-9].

Ever since Dai *et al.* [10] reported the initial manual attachment of a small multi-wall CNT (MWNT) onto a adhesive-coated AFM tip, tremendous efforts have been devoted to the development of such CNT-equipped SPM probes to leverage their high-aspect-ratios. Under a direct current (DC) electric field, pre-grown MWNTs were electrostatically oriented and mounted onto AFM probes by a number of research groups with the help of an electron microscope and a micro/nano-manipulator [11-14]. Alternatively, Lieber *et al.* [15] successfully picked up a single-wall CNT (SWNT) on an AFM tip while scanning across a substrate covered with pre-grown nanotube 'grass'. In parallel, direct growth from catalyst-loaded AFM probes by chemical vapor deposition (CVD) has been developed by the same group to improve reproducibility and scale up production [6, 16].

Under a heterogeneous alternating current (AC) electric field, the polarizability of CNTs has recently enabled their dielectrophoretic (DEP) assembly towards the high-field apex region of conductive SPM probes, rendering nanofibrils extended out from probe apex [17-19]. Simultaneous drawing from multiple STM needles was also achieved [20], demonstrating the parallel manufacturability of DEP assembly. This could provide an efficient and low-cost processing route for CNT-extended SPM probes, both of individual [19] or multiple nanotubes [20]. By varying the quantity of nanotubes constituting the nanofibril, CNT probes with various desirable length and diameters could be potentially obtained, enhancing their versatility in the diverse application environments forseen, with theoretically better control over probe stiffness at high-aspect-ratios as well due to its correlation to nanofibril geometry. For example, high-lateral-resolution tips with diameters as small as possible are desired in structural biology studies to elucidate fine features [4]. Robust high-aspect-ratio probes, on the other hand, are needed for lithographic photoresist patterns to characterize deep nanoscale trenches [5]. In localized electrochemistry [21, 22] within cellular environments [4], a nanoneedle with proper stiffness (depending on nanotube nanofibril dimension) is required in order to ensure successful cell membrane piercing [23]. We reported a comprehensive study in a previous contribution [24] on the DEP assembly process, including post-growth length-trimming and the growth of functional heterostructures. In this study, we are interested in the controlled tunability of nanofibril dimensions with the magnitude of the external applied AC electric field which is an important and controllable process parameter. In addition, the effects of tip geometry (cone angles and tip height) on the process of nanofibril dielectrophoretic growth are studied with a representative pool of eight different scanning probes, and their suitability for DEP assembly is evaluated.

**2. Experimental details**

Nitric acid (98%) and sulfuric acid (96.4%) were obtained from Sigma-Aldrich and used as is. A.C.S. reagent grade dimethylformamide (DMF) was purchased from J.T.Baker. Millipore quality deionized water (resistivity higher than 18MΩ) was used for all experiments. Laser ablated SWNTs were purchased from Tubes@Rice. Following the previously established protocol [25-30], pristine SWNTs were treated in a 3:1 mixture of $H_2SO_4$ and $HNO_3$ under sonication for 4 hours at 70°C, then filtered, washed with copious deionized water until the pH of filtrated water reached neutral, and finally dried overnight in vacuum. They were dispersed in DMF with the aid of sonication and further diluted in deionized water until nanotube suspension became colourless with a concentration of 0.003 mg/ml.

Commercial STM and AFM probes with various cone angles, tip heights, and force constants were used for carbon nanotube pulling to investigate probe geometry effects. The make/model information and technical specifications are listed in Table 1. Their representative optical images in side view are provided in Figure S1 of the Supporting Information. Since the tips are in various shapes including tetrahedron, cone and other geometries, the cone angle is defined as the planar angle measured from the optical images (Figure S1 of the Supporting Information) as the tips were observed from the side. Tip height is measured as the vertical distance from the tip end to the tip base where it attaches to the cantilever.

SWNT nanofibrils were attached onto the SPM probes using the positive dielectrophoresis technique. The schematic drawing of the experimental set-up is provided elsewhere [24, 31]. The scanning probe acted as the working electrode while a small hollow metal tube (outer diameter 600 μm, inner diameter 150 μm) was used as the counter electrode. The tip central axis was aligned parallel to that of the metal tube and a drop of SWNT dispersion was placed onto the end of the metal tube forming a droplet. An AC voltage of 2MHz, varied from 1-20V ($V_{pp}$, peak-to-peak) was applied between the two electrodes. In order to minimize evaporation of the SWNT solution, both electrodes were enclosed in a sealed environmental-cell (E-cell) covered with a thin transparent glass slide. On the top is an optical microscope equipped with a charge-coupled device (CCD) camera connected to a computer for digital monitoring and recording. The metal tube containing the nanotube solution was anchored on an XYZ 3D moving stage. The Y and Z axes were manually actuated providing alignment of the scanning probe apex with the summit of the solution droplet at the end of the metal tube. The electric-motor actuated X axis was automated for nanofibril drawing with steps down to 50 nm each, with the rate accelerated as drawing progresses.

The obtained SWNT nanofibrils were characterized with a high resolution field emission scanning electron microscope ('FESEM,' model JEOL 6335F) for morphological and dimensional measurements (length and diameter). The base pressure of the sample chamber was maintained at *c.a.* $10^{-6}$ torr, while the accelerating voltage was varied from 5-10 keV with the working distance set between 8-15 mm.

## 3. Results and Discussion

*3.1 Electric Field Effect on Dimension Control of Nanofibrils*

Dielectrophoresis (DEP) is the motion of a dielectric object as a result of polarization imparted by a non-uniform electric field. Upon application of an AC bias between the sharp conductive SPM probe and the relatively large metal tube counter-electrode, dipole moments are induced in the suspended SWNTs. The torque exerted on nanotubes leads their rotation to align their long axis parallel to the electric field direction [17, 19]. Besides torque, a force imbalance is present in the polarized nanotubes as a result of the anisotropy of the electric field. This non-zero net force causes a translational movement of SWNTs toward the high field region (the sharp SPM probe) if they are more polarizable with respect to the surrounding medium, which is termed positive dielectrophoresis. The mathematic expression of DEP force is given by equation (1) [19, 32, 33]:

$$F_{DEP} = \frac{2}{3}\pi d^2 l \varepsilon_m \text{Re}(K_A) \nabla |E|^2 \qquad (1)$$

$$K_A = \frac{\varepsilon_p^* - \varepsilon_m^*}{\varepsilon_m^*} \qquad (2)$$

$$\varepsilon^* = \varepsilon - i\frac{\sigma}{\omega} \qquad (3)$$

where $l$ and d are the length and diameter of carbon nanotube bundles, $\varepsilon_m$ is the dielectric constant of the medium, $\varepsilon_p$ is the dielectric constant of nanotubes, $\varepsilon^*$ is the complex permittivity expression, $Re(K_A)$ is the real part of Clausius-Mossotti factor $K_A$, $E$ is the electric field, $\sigma$ is the conductivity and $\omega$ is the frequency of the electric field. Besides forces imparted by the electric field, SWNTs are also subject to thermal Brownian motion, whose maximum value can be roughly given by:

$$F_{Thermal} = \frac{k_B T}{d} \qquad (4)$$

where $k_B$ is the Boltzmann constant, $T$ is the temperature. For $T$=300 K and $d$ = 10 nm, $F_{Thermal}$ is on the order of $4.14 \times 10^{-13}$ N. For SWNTs to migrate to the sharp AFM probes, DEP force must overcome this random force. If we take $l$ = 1µm, $\varepsilon_m$ = $80\varepsilon_0$ for deionized water (where $\varepsilon_0$ is vacuum permittivity), $Re(K_A) \approx 10^6$ for metallic SWNTs in water at 2 MHz [33] (metallic SWNTs dominate the dielectric constant of the bundles [24]), we have the gradient of electric field squared $\nabla|E|^2$ at least on the order of $2.8 \times 10^{12}$ from equation (1). $\nabla|E|^2$ is the combined contribution of electrode geometry and electric field intensity. As an approximation, our experimental setup can be treated as a typical cone-plane configuration. The mathematical description is thus given by the following relationship [32]:

$$\nabla|E|^2 = \nabla J^2 \frac{V_{RMS}^2}{S^3} \qquad (5)$$

where $V_{RMS}$ is the root mean square (RMS) voltage of external applied potential ($V_{RMS}=0.7V_{pp}/2$), $S$ is the spacing between SPM probes and the metal tube, $J$ is the field enhancement factor (FEF).

Take $V_{rms}$=3.5V ($V_{pp}$ = 10V, typical voltage for CNT AFM probe assembly) and $S$ = 300 µm, then the least gradient of field enhancement factor squared $\nabla|J|^2$ has to be larger than 6.2 for nanotube fibril assembly. Its magnitude increases rapidly with the decreasing distance to probe apex. Since electrode geometry is typically kept unchanged during nanofibril assembly, the question is that if the easily accessible potential parameter ($V_{pp}$) can be effectively varied to produce probes with tailored dimensions, which is the focus of this work we will present in the following.

Figure 1a illustrates representative results for the SWNT nanofibril AFM probes assembled under an AC electric field of 2MHz and 10V. The SWNT nanofibril exhibits a conical shape that is *c.a.* 5 μm long. The diameter at the connection point with the AFM probe is on the order of 100 nm, gradually narrowing down to 20-30 nm at the tip apex as a result of the accelerated translation process during tip drawing.

Figure 2 depicts the nanofibril length and diameter as a function of external applied potentials with all other conditions kept the same during tip pulling (frequency, tip geometry, drawing acceleration profile, etc.) in order to isolate the effect of electric field strength on nanofibril dimensions. Since the resulting nanofibrils were conical in geometry, the nanofibril diameter was measured at the position of the AFM tip apex. As the voltage ($V_{pp}$) increased in the range of 5-18V, both the nanofibril length and diameter enlarged asymptotically, with the averages increasing from 0.8μm and 30nm to 5μm and 100nm (Figure 1), respectively. Between voltages of 10-18V, this increase slowed significantly and entered into a relatively invariable regime. Specifically, for very low voltages (<4V), stable attachment of SWNT nanofibrils could not be achieved, presumably because DEP forces were insufficient. At very high voltages (≥20V), on the other hand, nanofibrils with very rough sidewalls were observed in FESEM images, suggesting the attachment of large amounts of unwanted debris from the SWNT suspension in addition to the nanotube bundles themselves. These points are discussed further below.

The nanofibril dimension is determined by the amount of nanotubes that precipitate out of the solution during the dielectrophoretic assembly, which is a direct result of their migration rate and hence the electric field parameters. Moving nanotubes in solution are subject to viscous resistance induced by friction between SWNTs and the surrounding medium. Assuming without fluid motion, the terminal velocity ($v_t$) which can be mathematically given as [33, 32]:

$$v_t = \left(\frac{F_{DEP}}{f}\right) \quad (6)$$

$$f = \frac{3\pi \eta \, l}{\ln(2l/d)} \quad (7)$$

where $f$ is the friction factor determined by the nanotube geometry and fluid viscosity ($\eta$). Take water viscosity $\eta = 1.0 \times 10^{-3}$ Pa·s, $l = 1$ μm and $d = 10$ nm, friction factor $f$ of $ca.$ $1.8 \times 10^{-9}$ is obtained, which gives a terminal speed of 230 μm/s under the least required DEP force. This is higher than our experimentally observed pulling speed (<100 μm/s) to halt nanofibril growth ref. This discrepancy originates from the lack of consideration of random forces present in the solution bubble (thermal Brown motion and possible fluid motion) in the calculation with equation (6).

The variation of DEP force changes the terminal speed and thus the dimension of grown nanofibrils. As indicated in equations (1) and (5), the DEP force is proportional to the external applied potential squared. As the applied voltage keeps increasing to 5-18V, the increasing DEP force exerted on nanotubes causes more nanotubes migrate and align, at the SPM probe apex as a result of increased velocities according to equation (6). This leads to an increased nanofibril dimension, and improved alignment along its axis. In the range of 5-10V, the dimension increase is nearly linear, in agreement with previously reported work at low fields [32]. However, the increase of nanofibril dimensions gradually reaches a plateau between 10-18V. This is because the migration rate of nanotubes is not just proportional to the DEP force according to voltages, but is also a function of friction forces produced in the liquid that eventually reach a maximum. More generally, for any given voltage it is important to note that as more nanotube bundles converge toward the probe apex where the field strength is highest, coalescence between individual bundles likely occurs resulting in an enlargement of bundle length and/or radius. This always increases the friction factor according to equation (7), preventing the terminal velocity from increasing limitlessly near the probe apex and thus conveniently ensuring stable and uniform nanofibril growth.

With the applied voltage further increased to 20V and beyond, debris in the solution started to attach along with the grown nanofibrils. Although the SWNT solution is purified, small impurities, including amorphous carbon and metal catalyst particles, are certainly present. Since the length of one-dimensional nanotubes is much larger than the size of impurities that are usually in irregular or near spherical shapes, the resulting DEP force is significantly higher for nanotubes. Lower electric fields (< 20V) are therefore not sufficient to induce the controlled migration of this small debris, which is also more susceptible to the randomly oriented Brownian motion. As voltage is increased, however, the DEP force induced by the electric field becomes able to overcome the Brownian motion and precipitate a large amount of these impurities along the carbon nanotube nanofibrils.

At very low voltages (<4V), DEP force could drop one or more order of magnitude. Take $V_{pp}$ = 3V ($V_{RMS}$ = 1.05V), the DEP force is less than one-tenth of that at the typical assembly voltage of $V_{pp}$ = 10V ($V_{RMS}$ = 3.5V). This implies the DEP force at low voltage could decrease to the order similar to or less than Brownian motion ($4.14 \times 10^{-13}$ N). This experimental results also confirmed the previous simulation results on DEP force on the order of $10^{-14}$-$10^{-10}$ N [33] for a similar setup. Although the DEF force could still be high enough to migrate nanotubes in the proximity of the SPM probe apex due to larger field enhancement factor, the amount of the attracted CNTs is not sufficient to form the protruding tips, instead they would coat on the probe surface.

To briefly summarize this section, the external applied field can be varied to produce probes with tailored dimensions for a wide range of diverse applications. The smaller dimensions of probes obtained with lower voltages (Figure 1b) are suitable for high-resolution imaging. The tip radius can even be further reduced if the initial nanotubes have smaller bundle sizes or are individually dispersed with the treatment of surfactants. Even so, typical high-aspect-ratio SWNT probes (Figure 1a) are routinely grown under 2MHz AC voltages of 10V, which is approximately the asymptotic dimension according to Figure 2. Higher voltage should be avoided because they have little effect on the nanofibril dimensions whereas the nanofibril quality degrades due to the incorporation of impurities.

*3.2 Tip Geometry Effect on Dielectrophoretic Assembly of Nanofibrils*

The diversity of commercially available scanning probes necessitates the investigation of the effects of initial tip geometry, surface hydrophilicity, and electrical conductivity on the dielectrophoretic growth of nanotube nanofibrils. In this part of our study, eight types of AFM and STM probes were employed to particularly assess the relevance of tip geometry (cone angles and height) for optimal nanofibril assembly. Technical specifications and respresentative images for these probes are provided in Table 1 and Figure S1, respectively. For each SWNT nanofibril assembly process, the same conditions were employed on all tips (10V, 2MHz) with results displayed in Figure 3.

Stable dielectrophoretic growth of nanotube nanofibrils was achieved on all probes except tips #5 and #8. Failure for these probes is attributed to their very small height or very large cone angle, which lead to an inability to form a properly controlled and stable meniscus upon tip immersion into the nanotube solution drop. Indeed, *in situ* optical measurements during nanofibril pulling with these probes revealed either the entire cantilever assembly (tip #5) or the full needle base (tip #8) was wetted, resulting in excessive capillary forces which disabled dielectrophoretic assembly. The tips that did support stable nanofibril assembly, on the other hand, represent a pool of typical commercially available SPM probes, with cone angles of 20-50 degrees and tip heights of 15-25 μm.

AFM probes are fabricated of Si or $Si_3N_4$ and STM needles are often composed of Pt/Ir or W (Table 1), which are hydrophilic on the surface. When in contact with the nanotube suspension drop, the attraction of the sharp tip for the liquid draws the liquid up onto the probe surface. Due to capillarity, the free-liquid surface assumes a shape with a concave meniscus (see image and schematic in Figure 4) with a contact angle of 50-60 degrees for Si or $Si_3N_4$ probes comparable to the reported values [34]. The resulting capillary force ($f_c$) is directly proportional to the surface tension ($\sigma$) and perimeter ($\pi d$), which is balanced by the contraction force ($F_c$) caused by the nanotube suspension drop (gray dashed lines in Figure 4) that is also the product of surface tension ($\sigma$) and corresponding perimeter ($\pi D$).

For similar surface hydrophilicity (thus similar contact angles), the length-scale dependent capillary force ($f_c$) will then increase as tip cone angles become larger since a larger diameter ($d$) occurs as a result of more tip wetting upon immersion (Figure 4 and Figure S2 of the Supporting Information). The corresponding balancing force ($F_c$) will also increase due to larger diameter ($D$). The tips # 1 and #2 in Figure 4 are two representative probes (standard and high-aspect-ratio respectively). They have dramatically different cone angles and the resulting immersion diameters $d$ are $ca$. 1 and 7 μm respectively. Take water surface tension $\sigma = 7.28 \times 10^{-2}$ N/m, the capillary force ($f_c$) is then estimated on the order of $2.3 \times 10^{-7}$ N and $1.6 \times 10^{-6}$ N respectively for tips #1 and #2, which is much larger in comparison with the DEP force. However, capillary force is competing with the adhesion force between carbon nanotubes and probe surfaces. Upon tip-drop separation, the capillary force tends to sweep the grown nanofibrils off the tip surface, with such effects becoming stronger as the tip cone angle increases. This is limiting for attaching nanofibrils to large-angle AFM probes, but beneficial for sharper tips as it has the result of essentially auto-aligning any SWNT bundles that are not initially aligned along the tip's axis.

As discussed in the forgoing analysis, the cone angle of tip #8 (94 degrees) was so large that excessive wetting (the immersion diameter $d$ is at least several hundred micrometers) occurred resulting in a tremendous capillary force. Since the tip surface and attached SWNTs are wet during the pulling process, there are sufficient small liquid molecules present between nanotube and tip surface, dramatically weakening their interactions (The important role of removing these residual molecules to enhance the tip nanotube bounding and nanotube-nanotube bounding was reported in our previous contribution.) [24]. Thus the capillary force is well beyond the adhesion force of nanofibrils onto tip surface. Another adverse effect of largely diverged tips is that they may result in a dramatically defocused electric field resulting in a reduced dielectrophoretic force in contrast to their sharp counterparts [35]. As a result, nanofibrils could not be assembled dielectrophoretically. Moreover, the excessive capillary force often causes undesired bending of the resulted nanofibrils, even producing nanohooks, due to the build-up of larger tension within nanofibrils.

Tip length is also an important parameter for controlled nanofibril pulling. Upon tip immersion into a solution drop, liquid climbs up onto the tip and forms a meniscus according to the cone angle and surface hydrophilicity. The typical wetting height is *c.a.* 5-7 μm for tips with cone angles of 30-50 degrees. If the tip is too short, such as tip #5 (height is only 5 μm), a controlled meniscus could not be formed during tip immersion; instead, the entire cantilever assembly behind the tip became wetted. Upon tip drop separation, the excessive capillary force then led to failure of the dielectrophoretic assembly.

Fabricating nanofibrils onto shorter probes is still possible if they have sufficiently narrow cone angles. As indicated in Figure 4 and Figure S2 of the Supporting Information, tips with narrow cone angles have a much smaller tip-drop interaction volume, which is defined as an envelope of liquid around the immersed tip portion. Smaller interaction volumes have inherently less volume to store the arriving nanotubes during DEP, which then precipitate out of solution onto their predecessors yielding shorter and narrower nanofibrils. This is consistent with experimental results of much shorter nanofibrils on tip #1 than those on tips #2, #3 and #4 (2-3μm *versus* 4-6μm in length). This characteristic provides a viable mechanism to tailor nanofibril length for different applications.

Based on the forgoing discussions, it is concluded that sharp tips with long shanks and small cone angles provide more predictable results and therefore enhanced process control. This is attributed to the reduced capillary force during SWNT tip pulling in combination with auto alignment of the nanofibril, stronger field focusing, and a smaller interaction volume. Experimentally, this result is confirmed, with such tips producing nanofibrils with better morphology, axial alignment, and length/diameter control. Accordingly, a phenomenological envelope indicating the most suitable probes for optimal SWNT dielectrophoretic assembly is suggested by the dashed line in Figure 3. As the tip cone angles become larger, longer tips are also necessary. However, this border-line is non-linear, but more like on an asymptotic trend. There exists a cut-off cone angle beyond which excessive capillary forces, dramatically reduced electric fields, or both result in failure of the dielectrophoretic assembly. Of course tip wetting can be adjusted by varying surface energies as well, providing a separate means for successful tip pulling on tips identified as unsuitable in this work (e.g. smaller tip heights and larger cone angles could be accommodated by increasing their surface hydrophobicity to reduce wetting). Therefore the envelope could be pushed to shorter and broader tips, further expanding the range of suitable SPM probes for dielectrophoretic assembly of CNT nanofibrils.

Finally, the spring constants of AFM probes are also relevant for nanofibril pulling as they determine the stiffness of their cantilevers. Unlike rigid STM needles and stiff cantilevers (e.g. tip #2), soft cantilevers (tips #3, #4 and #5) exhibit considerable deformation under the exerted capillary force during the nanofibril pulling process. Take tips #2 and #4 as an example, their capillary forces are *ca.* $1.6 \times 10^{-6}$ N and $2.24 \times 10^{-6}$ N respectively. Thus the horizontal force components along pulling direction ($F_h$) are $1.3 \times 10^{-6}$ N and $1.9 \times 10^{-6}$ N respectively according to $F_h = F_{capillary} \cdot \cos(\alpha - \theta/2)$ (see Figure 4), where $\alpha = 55°$ is contact angle and $\theta$ is tip cone angle. Then we have displacement of 0.65 μm and 9.5 μm for tips #2 and #4 respectively according to Hooke's law $F = kx$ ($k$ is tip spring constant, $x$ is tip displacement), which is consistent with experimental observations. In addition to distorting the electric field at the apex, this large deformation is particularly problematic when the tip and solution separate. Essentially, the attractive bending of the cantilever builds up a tension force (akin to drawing a bow) which is suddenly released upon separation, resulting in a large transient force drop which disturbs the dielectrophoretic assembly. This need not always occur for low spring constant cantilevers, however; for example, cantilever #1 is even softer than levers #3 and #4, but because of its small cone angle the correspondingly small capillary force induces negligible cantilever bending (~2.3 µm) and results in a high quality nanofibril probe. In conclusion, the tip geometry, surface hydrophilicity, and cantilever spring constant are all crucial to controlling the wetting behaviour during dielectrophoretic assembly, and therefore the quality of the resulting probes.

## 4. Conclusion

Positive dielectrophoresis was utilized to organize SWNT nanofibrils onto various conductive SPM probes to produce high-aspect-ratio nanoneedles. The magnitude of the applied electric field was varied from 1-20V to investigate its effect on nanofibril dimensions. Both length and diameter grew asymptotically as the voltage increased from 5-18V. At voltages below 4V, stable attachment of SWNT nanofibrils could not be achieved due to insufficient dielectrophoretic forces to overcome Brownian motion. At voltages of 20V or beyond, large amounts of impurities were also assembled onto the nanofibrils. Eight types of SPM probes were selected to study tip geometry effects and their suitability in the process of nanofibril assembly. Tip wetting plays a crucial role in the assembly process, which is closely related to the tip geometry (tip height and cone angles). Probes with narrow cone angles and long shanks exhibit stronger field focusing and have smaller tip portions wetted, which incurs smaller capillary forces during tip pulling and therefore shorter and higher quality nanofibril probes. Applications for such nanofibril-decorated scanning probes are broad, including high resolution imaging of deep nanostructures [24], biomechanics studies, electrochemical analysis and scanning surface potential microscopy [31].

**Acknowledgements**

Financial support from U.S. Army Research Office (grant # ARO-DAAD-19-02-1-0381) is greatly appreciated.

**Supporting Information Available:** Optical images of SPM probes with various tip geometry and spring constants (Figure S1); Tip wetting behaviour of SPM probes with different cone angles (Figure S2) (PDF). This material is available free of charge via the Internet at http://www.iop.org.

**References**

[1] Bonnell D A 2001 *Scanning probe microscopy and spectroscopy: theory, techniques, and applications* (New York: Wiley-VCH)

[2] Frisbie C D, Rozsnyai L F, Noy A, Wrighton M S and Lieber C M 1994 Functional-Group Imaging by Chemical Force Microscopy *Science* **265** 2071-4

[3] Noy A, Frisbie C D, Rozsnyai L F, Wrighton M S and Lieber C M 1995 Chemical Force Microscopy - Exploiting Chemically-Modified Tips to Quantify Adhesion, Friction, and Functional-Group Distributions in Molecular Assemblies *J. Am. Chem. Soc.* **117** 7943-51

[4] Hafner J H, Cheung C L, Woolley A T and Lieber C M 2001 Structural and functional imaging with carbon nanotube AFM probes *Prog. Biophys. Mol. Biol.* **77** 73-110

[5] Nguyen C V, Stevens R M D, Barber J, Han J, Meyyappan M, Sanchez M I, Larson C and Hinsberg W D 2002 Carbon nanotube scanning probe for profiling of deep-ultraviolet and 193 nm photoresist patterns *Appl. Phys. Lett.* **81** 901-3

[6] Cheung C L, Hafner J H and Lieber C M 2000 Carbon nanotube atomic force microscopy tips: direct growth by chemical vapor deposition and application to high-resolution imaging *Proc. Natl. Acad. Sci. U. S. A.* **97** 3809-13.

[7] Nguyen C V, Chao K J, Stevens R M D, Delzeit L, Cassell A, Han J and Meyyappan M 2001 Carbon nanotube tip probes: stability and lateral resolution in scanning probe microscopy and application to surface science in semiconductors *Nanotechnol.* **12** 363-7

[8] Wong S S, Harper J D, Lansbury P T and Lieber C M 1998 Carbon nanotube tips: High-resolution probes for imaging biological systems *J. Am. Chem. Soc.* **120** 603-4

[9] Yu M-F, Files B S, Arepalli S and Ruoff R S 2000 Tensile loading of ropes of single wall carbon nanotubes and their mechanical properties *Phys. Rev. Lett.* **84** 5552-5

[10] Dai H, Hafner J H, Rinzler A G, Colbert D T and Smalley R E 1996 Nanotubes as nanoprobes in scanning probe microscopy *Nature* **384** 147-50

[11]     Nishijima H, Kamo S, Akita S, Nakayama Y, Hohmura K I, Yoshimura S H and Takeyasu K 1999 Carbon-nanotube tips for scanning probe microscopy: Preparation by a controlled process and observation of deoxyribonucleic acid *Appl. Phys. Lett.* **74** 4061-3

[12]     Stevens R, Nguyen C, Cassell A, Delzeit L, Meyyappan M and Han J 2000 Improved fabrication approach for carbon nanotube probe devices *Appl. Phys. Lett.* **77** 3453-5

[13]     Akita S, Nishijima H, Nakayama Y, Tokumasu F and Takeyasu K 1999 Carbon nanotube tips for a scanning probe microscope: their fabrication and properties *J. Phys. D: Appl. Phys.* **32** 1044-8

[14]     Martinez J, Yuzvinsky T D, Fennimore A M, Zettl A, Garcia R and Bustamante C 2005 Length control and sharpening of atomic force microscope carbon nanotube tips assisted by an electron beam *Nanotechnol.* **16** 2493-6

[15]     Hafner J H, Cheung C L, Oosterkamp T H and Lieber C M 2001 High-yield assembly of individual single-walled carbon nanotube tips for scanning probe microscopies *J. Phys. Chem. B* **105** 743-6

[16]     Hafner J H, Cheung C L and Lieber C M 1999 Growth of nanotubes for probe microscopy tips *Nature* **398** 761-2

[17]     Zhang J, Tang J, Yang G, Qiu Q, Qin L-C and Zhou O 2004 Efficient fabrication of carbon nanotube point electron sources by dielectrophoresis *Adv. Mater.* **16** 1219-22

[18]     Tang J, Yang G, Zhang Q, Parhat A, Maynor B, Liu J, Qin L-C and Zhou O 2005 Rapid and reproducible fabrication of carbon nanotube AFM probes by dielectrophoresis *Nano Lett.* **5** 11-4

[19]     Lee H W, Kim S H, Kwak Y K and Han C S 2005 Nanoscale fabrication of a single multiwalled carbon nanotube attached atomic force microscope tip using an electric field *Rev. Sci. Instrum.* **76** 046108_1-_5

[20] Tang J, Gao B, Geng H, Velev O D, Qin L-C and Zhou O 2003 Assembly of 1D nanostructures into sub-micrometer diameter fibrils with controlled and variable length by dielectrophoresis *Adv. Mater.* **15** 1352-5

[21] Boo H, Jeong R A, Park S, Kim K S, An K H, Lee Y H, Han J H, Kim H C and Chung T D 2006 Electrochemical nanoneedle biosensor based on multiwall carbon nanotube *Anal. Chem.* **78** 617-20

[22] Campbell J K, Sun L and Crooks R M 1999 Electrochemistry using single carbon nanotubes *J. Am. Chem. Soc.* **121** 3779-80

[23] Obataya I, Nakamura C, Han S, Nakamura N and Miyake J 2005 Nanoscale Operation of a Living Cell Using an Atomic Force Microscope with a Nanoneedle *Nano Lett.* **5** 27-30

[24] Wei H, Kim S N, Zhao M, Ju S-Y, Huey B D, Marcus H L and Papadimitrakopoulos F 2008 Control of Length and Spatial Functionality of Single-Wall Carbon Nanotube AFM Nanoprobes *Chem. Mater.* **20** 2793-801

[25] Liu J, Rinzler A G, Dai H, Hafner J H, Bradley R K, Boul P J, Lu A, Iverson T, Shelimov K, Huffman C B, Rodriguez-Macias F, Shon Y-S, Lee T R, Colbert D T and Smalley R E 1998 Fullerene pipes *Science* **280** 1253-6

[26] Liu J, Casavant M J, Cox M, Walters D A, Boul P, Lu W, Rimberg A J, Smith K A, Colbert D T and Smalley R E 1999 Controlled deposition of individual single-walled carbon nanotubes on chemically functionalized templates *Chem. Phys. Lett.* **303** 125-9

[27] Chattopadhyay D, Galeska I and Papadimitrakopoulos F 2002 Complete elimination of metal catalysts from single wall carbon nanotubes *Carbon* **40** 985-8

[28] Wei H, Kim S N, Marcus H L and Papadimitrakopoulos F 2006 Preferential forest assembly of single-wall carbon nanotubes on low-energy electron-beam patterned nafion films *Chem. Mater.* **18** 1100-6

[29]  Wei H, Kim S, Kim S N, Huey B D, Papadimitrakopoulos F and Marcus H L 2007 Patterned Forest-Assembly of Single-Wall Carbon Nanotubes on Gold Using a Non-Thiol Functionalization Technique *J. Mater. Chem.* **17** 4577-85

[30]  Wei H, Kim S N, Kim S, Huey B D, Papadimitrakopoulos F and Marcus H L 2008 Site-specific Forest-assembly of Single-Wall Carbon Nanotubes on Electron-beam Patterned $SiO_x$/Si Substrates *Mater. Sci. Eng., C* DOI: 10.1016/j.msec.2008.03.002

[31]  Zhao M, Sharma V, Wei H, Birge R R, Stuart J A, Papadimitrakopoulos F and Huey B D 2008 Ultrasharp and high aspect ratio carbon nanotube atomic force microscopy probes for enhanced surface potential imaging *Nanotechnol.* **19** 235704

[32]  Hulman M and Tajmar M 2007 The dielectrophoretic attachment of nanotube fibres on tungsten needles *Nanotechnol.* **18**

[33]  Kim J E and Han C S 2005 Use of dielectrophoresis in the fabrication of an atomic force microscope tip with a carbon nanotube: a numerical analysis *Nanotechnol.* **16** 2245-50

[34]  Tao Z H and Bhushan B 2006 Wetting properties of AFM probes by means of contact angle measurement *Journal of Physics D-Applied Physics* **39** 3858-62

[35]  Sacha G M, Verdaguer A, Martinez J, Saenz J J, Ogletree D F and Salmeron M 2005 Effective tip radius in electrostatic force microscopy *Appl. Phys. Lett.* **86** 123101

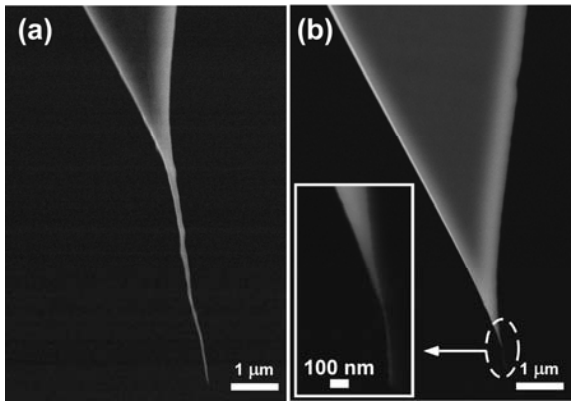
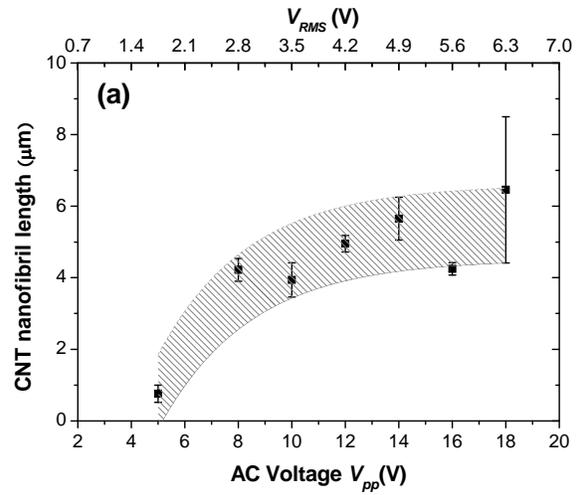

**Figure 1.** FESEM micrographs of typical high-aspect-ratio SWNT AFM nanoneedles grown dielectrophoretically under 2MHz AC biases of 10V (a) and 5V (b), respectively.

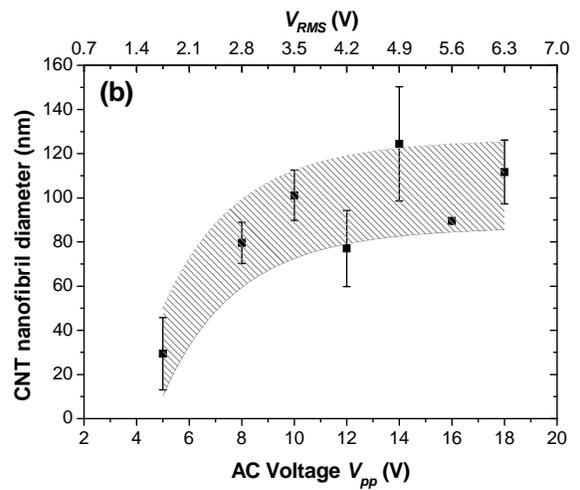

**Figure 2.** The effects of the magnitude of an AC electric field on SWNT nanofibril length (a) and diameter (b).

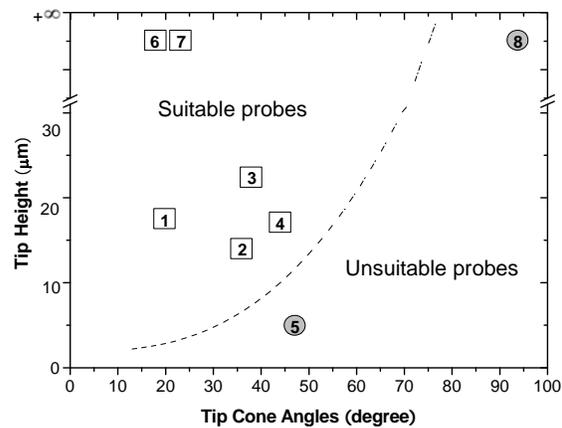 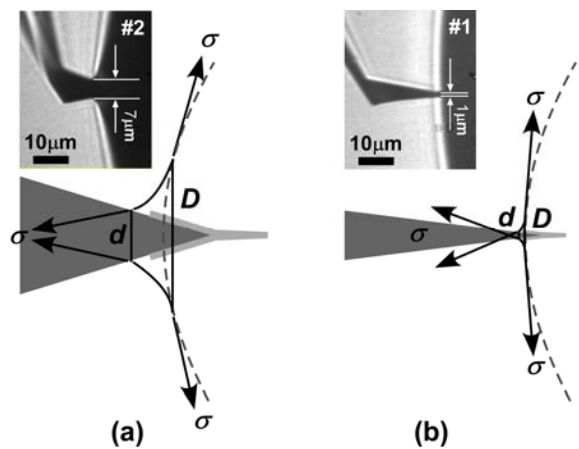

**Figure 3.** The dielectrophoretic assembly of SWNT nanofibrils on SPM probes with various tip heights and cone angles. White squares indicate a stable successful assembly. Gray circles denote an unsuccessful growth. The numbers inside are the tip indices as listed in Table 1. The dashed line indicates the suitable range of probes for DEP assembly.

**Figure 4.** Schematic representation of meniscus formation upon immersion of SPM probes in the nanotube suspension drop and the resulting capillary force (see text for details).

**Table 1.** Technical specifications of the selected SPM probes.

| SPM probes | No. | Make_Model (Probe materials) | Cone angle (degrees) | Tip height (μm) | Force constant (N/m) |
|---|---|---|---|---|---|
| AFM tips | 1 | Nanosensors ATEC-CONT (Si) | 20 | 15-20 | 0.061-0.082[a] |
|  | 2 | Olympus AC240 (Si) | 36 | 14 | 2 |
|  | 3 | μmasch CSC38/Ti-Pt (Si coated with Ti-Pt) | 38 | 20-25 | 0.234-0.297[a] |
|  | 4 | Budget Sensors BS-Cont E (Si) | 44 | 17 | 0.2 |
|  | 5 | Biolever ($Si_3N_4$) | 47 | 5 | 0.02 |
| STM needles | 6 | Digital Instrument (Pt-Ir) | 19 | - | - |
|  | 7 | Angstrom Technology (W) | 22 | - | - |
|  | 8 | Molecular Imaging (Pt-Ir) | 94 | - | - |

[a] measured values

# Supporting information

# Electric Field and Tip Geometry Effects on Dielectrophoretic Growth of Carbon Nanotube Nanofibrils on Scanning Probes

*Haoyan Wei[1,3], Anna Craig[1], Bryan D. Huey[1], Fotios Papadimitrakopoulos[2] and Harris L. Marcus[1,4]*

[1] Materials Science and Engineering Program, Department of Chemical, Materials and Biomolecular Engineering, Institute of Materials Science, University of Connecticut, Storrs, CT 06269, USA

[2] Nanomaterials Optoelectronics Laboratory, Polymer Program, Institute of Materials Science, Department of Chemistry, University of Connecticut, Storrs, CT 06269, USA

[3] Present address: Applied Sciences Laboratory, Institute for Shock Physics, Washington State University, Spokane, WA 99210, USA, email: hwei@wsu.edu

[4] To whom correspondence should be addressed.

E-mail: hmarcus@ims.uconn.edu     Phone: (860) 486-4623     Fax: (860) 486-4745

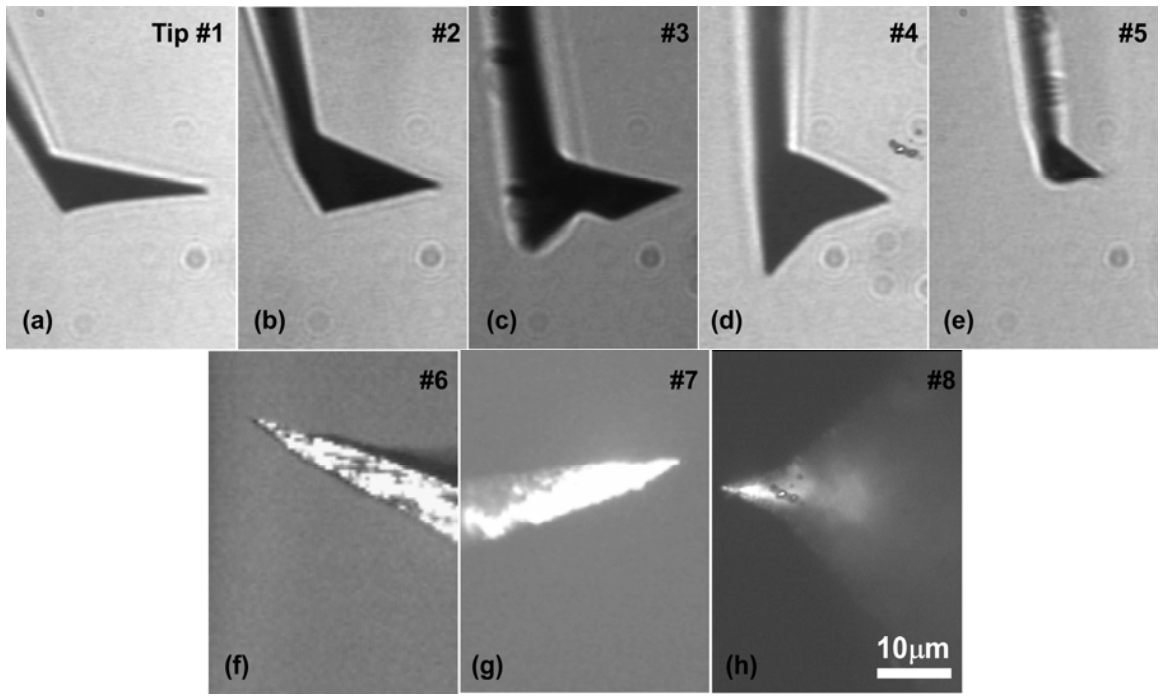

**Figure S1.** Representative optical images of SPM probes with various tip geometry and force constants (see Table 1 in the main text for their technical specifications).

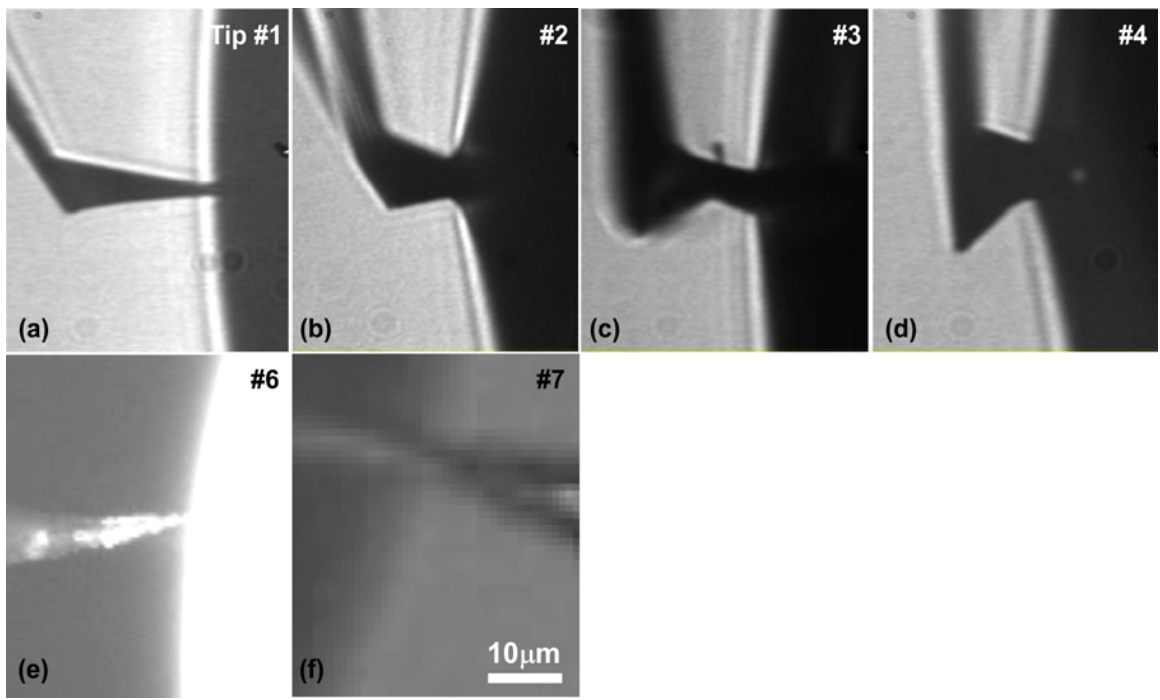

**Figure S2.** The immersion of SPM probes in nanotube suspension bubble. Reduced tip wetting is obtained on tips with smaller cone angles.